\documentclass[aps, prb,twocolumn,showpacs,preprintnumbers,amsmath,amssymb]{revtex4}

\usepackage{graphicx}% Include figure files
\usepackage{dcolumn}% Align table columns on decimal point
\usepackage{bm}

\begin{document}
\preprint{PRB/Y. K. Li et al.}

\title{Effect of Co doping on superconductivity and transport properties in SmFe$_{1-x}$Co$_{x}$AsO}
\author{Y. K. Li, X. Lin, Z. W. Zhu, H. Chen, C. Wang, L. J. Li, Y. K. Luo, M. He, Q. Tao, H. Y. Li, G. H. Cao\footnote[1]{Electronic address: ghcao@zju.edu.cn} and Z. A. Xu\footnote[2]{Electronic address: zhuan@zju.edu.cn}}

\affiliation{$^{1}$Department of Physics, Zhejiang University, Hangzhou 310027, People's Republic of China}

\date{\today}

\begin{abstract}
We investigate superconductivity and transport properties of Co
doped SmFe$_{1-x}$Co$_{x}$AsO system. The antiferromagnetic (AFM)
spin-density wave (SDW) order is rapidly suppressed by Co doping,
and superconductivity emerges as $x$ $\geq$ 0.05. $T_c$$^{mid}$
increases with increasing Co content, shows a maximum of 17.2 K at
the optimally doping of $x\sim$ 0.10. A phase diagram is derived
based on the transport measurements and a dome-like $T_c$ versus
$x$ curve is established. Meanwhile we found that the normal state
thermopower might consist of two different contributions. One
contribution increases gradually with increasing $x$, and the
other contribution is abnormally enhanced in the superconducting
window 0.05 $\leq$ $x$ $\leq$ 0.20, and shows a dome-like doping
dependence. A close correlation between $T_{c}$ and the abnormally
enhanced term of thermopower is proposed.

\end{abstract}
\pacs{74.70.Dd; 74.62.Dh; 74.25.Fy; 74.25.Dw}

\maketitle

\section{\label{sec:level1}Introduction}

Soon after the discovery of superconductivity at 26 K in
LaO$_{1-x}$F$_{x}$FeAs\cite{Hosono}, the substitution of La by
other rare earth elements such as Ce\cite{WNL-Ce},
Pr\cite{Ren-Pr}, Sm\cite{Chen-Sm,Ren}, Nd\cite{Ren},
Gd\cite{WenGd,Wang-Th}, and Tb\cite{BosTb,LiTb} has led to a
family of 1111 phase high-\emph{T$_c$} superconductors. The parent
compounds of the new 1111 phase materials, LnFeAsO (Ln = La, Ce,
Pr, Sm, Nd, Gd, and Tb etc.), have a quasi two-dimensional
tetragonal structure, consisting of insulating Ln$_2$O$_2$ layers
and conducting Fe$_2$As$_2$ layers. Similar to high-$T_c$
cuprates, superconductivity occurs through electron (or hole)
doping into an antiferromagentic (AFM) parent compound. In the
high-$T_c$ cuprates, superconductivity is induced by chemical
doping in "charge reservoir" layers which are out of the
superconducting CuO$_2$ planes. Meanwhile, the substitution of Cu
with other 3$d$ elements such as Ni and Zn in the CuO$_2$ planes
severely destroys the superconductivity\cite{Xiao,Tarascon}. In
contrast to high-$T_c$ cuprates, it has been found that
superconductivity can also be induced by partial substitution of
Fe by other transition metal elements in the
superconducting-active Fe$_2$As$_2$ layers. Sefat \emph{et
al.}\cite{SefatCo} reported first the superconductivity in the Co
doped LaFe$_{1-x}$Co$_x$FeAs. We also independently found that
both Co doping\cite{CaoCo} and Ni doping\cite{CaoNi} can induce
superconductivity in the LaOFeAs system. It is also found that the
doping of non-magnetic impurities Zn$^{2+}$ ions in the
Fe$_2$As$_2$ conducting layers affects selectively the AFM order,
and superconductivity remains almost unperturbed in
LaFe$_{1-x}$Zn$_x$AsO$_{1-y}$F$_{y}$ system\cite{LiZn}. This
implies that superconductivity is quite robust to the disorder in
the conducting Fe$_2$As$_2$ layer, which might be taken for a
significant difference between the high-$T_c$ cuprates and
iron-based arsenide superconductors.

Compared to the phase diagram of F-doped
LaFeAsO\cite{Hosono,wnl,PD}, the phase diagram of Co-doped LaFeAsO
system shows some significant differences\cite{CaoCo}. Firstly, Co
doping destroys the AFM SDW order more strongly. Secondly, the
maximum $T_{c}$ is significantly lowered in Co-doped system.
Finally the optimal doping level is distinctly lower and the
superconducting window is much narrower in Co doped
LaFe$_{1-x}$Co$_{x}$AsO system. The disorder effect caused by Co
doping within (Fe/Co)As layers can not be ignored.

In this paper, we investigate in detail the Co doping effect on
the superconductivity and transport properties of
SmFe$_{1-x}$Co$_{x}$AsO, and an electronic phase diagram is
derived. A dome-like Co doping ($x$) dependence of $T_c$ is
established. Furthermore, we have found that the normal state
thermopower increases remarkably with Co doping in the
"underdoped" region. A possible correlation between $T_c$ and the
enhanced thermopower in the superconducting window (0.05 $\leq$
$x$ $\leq$ 0.20) is proposed.

\section{\label{sec:level1}Experimental}

The polycrystalline samples of SmFe$_{1-x}$Co$_{x}$AsO were
prepared in vacuum by solid state reaction using SmAs,
Sm$_2$O$_3$, Fe$_2$As, FeAs, and Co$_3$O$_4$ as starting
materials. SmAs was pre-synthesized by reacting Sm slices and As
powders at 1173 K for 24 hours in an evacuated quartz tube. FeAs
and Fe$_{2}$As were obtained by reacting the mixture of
stoichiometric element powders at 873 K for 10 hours,
respectively. Co$_{3}$O$_{4}$ and La$_{2}$O$_{3}$ were dried in
air at 773 K and 1173 K, respectively, for 24 hours before using.
Then all powders of these intermediate materials were accurately
weighed according to the stoichiometric ratio of
SmFe$_{1-x}$Co$_{x}$AsO ($x$=0, 0.01, 0.025, 0.05, 0.075, 0.1,
0.125, 0.15, 0.175, 0.2, 0.225, 0.25 and 0.3), thoroughly mixed in
an agate mortar, and pressed into pellets under a pressure of 2000
kg/cm$^{2}$. All the processes were operated in a glove box filled
with high-purity argon. Finally the pellets were sintered in an
evacuated quartz tube at 1423 K for 40 hours and furnace-cooled to
room temperature.

Powder X-ray diffraction (XRD) was performed at room temperature
using a D/Max-rA diffractometer with Cu K$_{\alpha}$ radiation and
a graphite monochromator. Lattice parameters were calculated by a
least-square fit using at least 20 XRD peaks in the range of
$20^{\circ}\leq 2\theta \leq 80^{\circ}$. The errors were
estimated as three times of the standard deviations of the fit.
The electrical resistivity was measured by four-terminal method.
The temperature dependence of d.c. magnetization was measured on a
Quantum Design magnetic property measurement system (MPMS-5) with
an applied field of 10 Oe. The thermopower was measured by a
steady-state technique, and the applied temperature gradient was
less than 0.5 K/mm.

\section{\label{sec:level1}Results and Discussion}

Fig.1(a) shows the representative XRD patterns of the
SmFe$_{1-x}$Co$_{x}$AsO samples. The diffraction peaks of all the
samples can be well indexed based on a tetragonal cell of
ZrCuSiAs-type structure, which indicates that the samples are all
pure phase. Fig.1(b) shows the variations of refined lattice
parameters with Co content($x$). Co doping causes the shrinkage of
$c$-axis significantly, while the $a$-axis remains nearly
unchanged. Thus the cell volume decreases monotonously with $x$,
which is related to the smaller Co$^{2+}$ ions (than Fe$^{2+}$
ions). This fact indicates that Co is successfully doped into the
lattice, according to Vegard's law. The shrinkage of $c$-axis
suggests the strengthening of interlayer Coulomb attraction,
implying the increase of density of negative charge in
Fe$_2$As$_2$ layers by the Co doping. Similar variations of
lattice constants were also observed in the Co doped
LaFe$_{1-x}$Co$_{x}$AsO in the previous report\cite{CaoCo}.

\begin{figure}
\includegraphics[width=8cm]{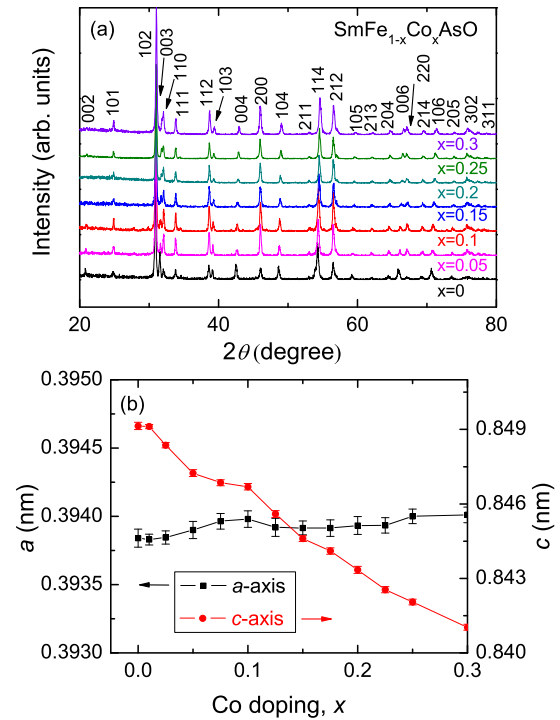}
\caption{(Color online) Structural characterization of
SmFe$_{1-x}$Co$_{x}$AsO samples. (a) Powder X-ray diffraction
patterns of representative SmFe$_{1-x}$Co$_{x}$AsO samples. (b)
Lattice parameters as a function of Co content.}
\end{figure}

Fig.2 shows the temperature dependence of electrical resistivity
($\rho$) of SmFe$_{1-x}$Co$_{x}$AsO samples in the temperature
range from 3 K to 300 K. The inset shows an enlarged plot of
$\rho$ versus $T$ for the low temperatures. For the undoped parent
compound, a clear drop in the resistivity is observed below about
140 K just as in the case of LnOFeAs\cite{Hosono}, which has been
ascribed to a structural phase transition and antiferromagnetic
spin-density-wave (SDW) transition\cite{DaiPC,Dai2}. This
anomalous temperature $T_{an}$, which is defined as the peak
position in the temperature dependence of the derivative of
resistivity, decreases from 137 K for $x$ =0, to 124 K and 93 K
for $x$ = 0.01 and 0.025, respectively. For $x$ = 0.05, such an
anomalous change in resistivity almost disappears, and only a tiny
kink around 45 K can be distinguished. Within the doping range
0.05 $\leq$ $x$ $\leq$ 0.20, superconducting transition can be
observed at low temperatures. Meanwhile, the resistivity anomaly
disappears completely for $x$ $>$ 0.05. This means that the
superconductivity occurs wherefrom the suppression of SDW order.
Superconducting transition temperature $T_{c}^{mid}$, which is
defined as the midpoint in the resistive transition, reaches a
maximum of 17.2 K at the "optimally doped" level $x$ = 0.1. This
maximum of $T_{c}^{mid}$ is larger than that of Co-doped
LaFe$_{1-x}$Co$_{x}$AsO system\cite{CaoCo}. The volume fraction of
magnetic shielding is over 60\% for the "optimally" doped sample
estimated according to its magnetic susceptibility (not shown
here). Furthermore, the "superconducting window" is in the doping
range 0.05 $\leq$ $x$ $\leq$ 0.20, which is also larger compared
to the superconducting window (0.025 $\leq x\leq $ 0.125) for
LaFe$_{1-x}$Co$_{x}$AsO system.

Another interesting feature is that there exists a resistivity
minimum at $T_{min}$ in the normal state in the underdoped and
overdoped regimes. The resistivity changes from metallic into
semiconductor-like as $T$ $<$ $T_{min}$, \emph{i.e.}, there exists
a crossover from metal into insulator as $T$ decreases. However,
such a resistivity upturn disappears in the doping regime 0.15
$\leq$ $x$ $\leq$ 0.175. We suggest that this upturn could be
hidden in the strong superconducting fluctuations as
$T_c^{onset}$, the onset point in the resistive transition, is
quite high in this regime. Actually, such a crossover persists to
$x$ $>$ 0.20 for Co doped LaFe$_{1-x}$Co$_{x}$AsO
system\cite{CaoCo}. Meanwhile the room temperature resistivity
shows a monotonous decrease with increasing $x$. In the region of
large Co doping level ($x$ $>$ 0.15), the temperature dependence
of resistivity follows a power law for temperature range $T$ $>$
$T_{min}$, \emph{i.e.}, $\rho$ $\propto$ $T^n$. The index $n$ is
about 1.65 for $x$ = 0.25. It is clear that the system becomes
more metallic with increasing Co content, consistent with the
itinerant character of Fe 3$d$ electrons in the iron-based
oxy-arsenides revealed by the band structure calculations and
theoretical analysis\cite{Singh,wnl,Tesanovic}. Furthermore, the
theoretic calculation\cite{Xu} reveals that total electron
density-of-states (DOS) for LaO$M$As ($M$ = Mn, Fe, Co and Ni)
remains basically unchanged, except that Fermi level shifts toward
the top of valence band with band filling (adding electrons) one
by one from $M$ = Mn, Fe, Co to Ni. According to this calculation,
partial substitution of iron by cobalt is expected to add
electrons into Fe$_2$As$_2$ layers, and thus the more metallic
state is expected with increasing $x$.

As noted in the previous report\cite{CaoCo}, the possibility that
oxygen deficiency itself might induce superconductivity in this
system can be excluded. By high-pressure synthesis,
superconductivity was indeed observed in oxygen-deficient
LnFeAsO$_{1-\delta}$\cite{Ren-Sm2,Kito}. It has also been reported
that superconductivity was induced by oxygen deficiency in
Sr-doped LaFeAsO via annealing in vaccum.\cite{Wu} We note that
all the reported superconductors showed a \emph{remarkable}
decrease in $a$-axis as well as $c$-axis owing to the oxygen
deficiency. In contrast, the present SmFe$_{1-x}$Co$_{x}$AsO
samples show no obvious change in the $a$-axis, suggesting no
significant oxygen deficiency.

\begin{figure}
\includegraphics[width=8cm]{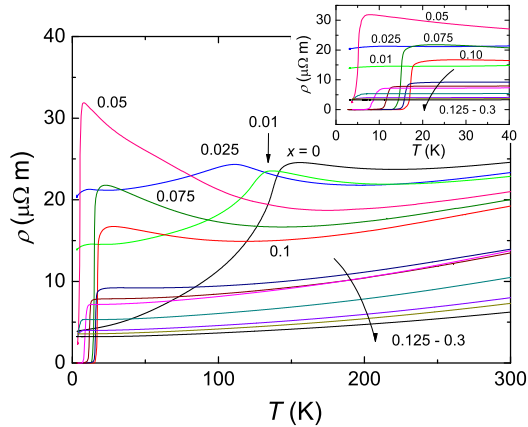}
\caption{(Color online) Temperature dependence of resistivity
($\rho$) for the SmFe$_{1-x}$Co$_{x}$AsO samples. Inset: the
enlarged plot of $\rho$ verus $T$ for low temperatures, showing
the superconducting transitions.}
\end{figure}

Based on above resistivity data, an electronic phase diagram for
SmFe$_{1-x}$Co$_{x}$AsO was thus established, as depicted in Fig.
3. The phase region of the SDW state is very narrow. 5\% Co doping
completely destroys the SDW order, and superconductivity emerges.
In the superconducting window (0.05 $\leq$ $x$ $\leq$ 0.20), a
dome-like $T_c(x)$ curve is observed, similar to that of cuprate
superconductors. Similar dome-like doping denpendence of $T_c$ is
also established for LaFe$_{1-x}$Co$_{x}$AsO\cite{CaoCo}. However,
the superconducting window of SmFe$_{1-x}$Co$_{x}$AsO system is
larger compared to that of LaFe$_{1-x}$Co$_{x}$AsO system. Though
the normal state shows metallic conduction at high temperatures,
the upturn in resistivity is observed at low temperatures in a
large doping region. For the higher Co-doping levels ($x\geq$
0.20), superconductivity no longer survives, but the resistivity
becomes more metallic. Complete replacement of Fe by Co is
possible, but whether SmCoAsO is an itinerant ferromagnetic metal
like LaCoAsO \cite{LaCoAsO} need to be clarified.

\begin{figure}
\includegraphics[width=8cm]{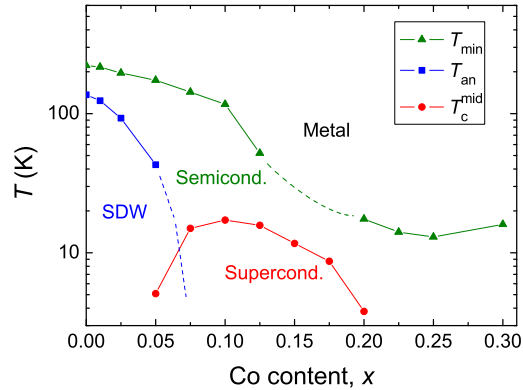}
\caption {(Color online) The electronic phase diagram for
SmFe$_{1-x}$Co$_{x}$AsO. $T_{min}$ separates the metallic and
semiconducting regions in the normal state of the superconductors.
Note that the vertical axis is in logarithmic scale.}
\end{figure}

Fig. 4 shows the temperature dependence of normal state
thermopower ($S$) for SmFe$_{1-x}$Co$_{x}$AsO samples. All of the
thermopowers are negative in the whole temperature range, which
means that the electron-like charge carriers dominate. For the
undoped parent compound, thermopower starts to increase abnormally
around $T_{an}$ at which the resistivity starts to decrease.
Similar anomalous increase in the thermopower below $T_{an}$ has
been reported in the undoped parent compounds
LaFeAsO\cite{McGuire} and TbFeAsO\cite{LiTb}. Such a remarkable
change in the thermopower should be caused by the change in the
electronic state when the system undergoes the structural phase
transition and SDW transition. This anomaly is gradually
suppressed with increasing Co doping, and disappears for $x$ $>$
0.05, consistent with the resistivity data. For the
superconducting samples, the profile of $S(T)$ curves is very
similar to that of high-$T_c$ cuprates except that it is negative
for SmFe$_{1-x}$Co$_{x}$AsO sytem. However, in contrast to
high-$T_c$ cuprates where the value of normal state thermopower
decreases monotonously with increasing doping
level\cite{Obertelli,Tallon}, the absolute value of thermopower,
$|S|$, increases quickly with Co doping, and the maximum in $|S|$
is about 80 $\mu$V/K for optimally doped level ($x$ = 0.1). Such a
large value of $|S|$ is very unusual in superconducting materials.
However, the remarkable enhancement of $|S|$ has also been
observed in F doped
LaFeAsO$_{1-x}$F$_x$\cite{Dragoe,Sefat,McGuire} and in Th doped
Tb$_{1-x}$Th$_x$FeAsO\cite{LiTb}. It should be a universal feature
for iron-based arsenide superconductors. It has also been proposed
that the F-doped iron-oxypnictides can be promising thermoelectric
materials in refrigeration applications around liquid nitrogen
temperatures \cite{Dragoe}. A rough estimate of $|S|$ according to
the Mott expression gives a value of less than 10 $\mu$V/K for F
doped LnFeAsO \cite{Sefat}. Whether the enhanced thermopower is
associated with strong electron correlation, magnetic
fluctuations, or specific electronic structure is an open issue.

\begin{figure}
\includegraphics[width=8cm]{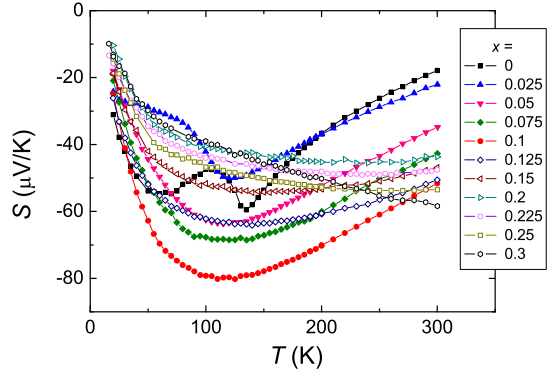}
\caption {(Color online) Temperature  dependence of thermopower
($S$) for SmFe$_{1-x}$Co$_{x}$AsO samples.}
\end{figure}

It is well established that there is a universal doping (hole
concentration) dependence of superconducting transition
temperature, $T_c$, for high-$T_c$ cuprates. Furthermore, it has
been found that there exists a close correlation between the room
temperature thermopower, $S$(290K), and the hole concentration,
$p$, and thus a universal correlation between $T_c$ and $S$(290K)
is observed\cite{Obertelli,Tallon}. In order to explore the
possible relationship between thermopower and superconducting
transition temperature in this system, we also plot both $S$(300K)
and $T_{c}^{mid}$ versus the doping level ($x$) for
SmFe$_{1-x}$Co$_{x}$AsO sytem. It becomes obvious that $S$(300K)
increases with $x$ as $T_{c}^{mid}$ does for $x$ $<$ 0.1, reaches
a maximum at $x$ = 0.1, and then gradually decreases with $x$ in
the overdoped region. For $x$ $>$ 0.2, superconductivity
disappears and the thermopower starts to increase again. Actually
it can be seen from Fig.5, that there seems to be two different
contributions to the thermopower, \emph{i.e.}, $S$(300K) =
$S_0$(300K) + $S'$(300K). The first term $S_0$(300K) is the normal
contribution (shown by the dashed line in the superconducting
window), which increase gradually with increasing $x$. The other
term $S'$(300K) only appears in the superocnducting window (shown
by the blue open symbols in Fig.5), which shows a dome-like doping
dependence as $T_{c}^{mid}$ does. We propose that there should be
a close correlation between superconducting state and the
anomalous term $S'$(300K). It will be an interesting issue whether
such a correlation between $T_c$ and $S'$(300K) is a universal
feature for all the iron-based arsenide superconductors.

The anomalous contribution to the thermopower, represented by
$|S'|$(300K), is hard to understand in frame of conventional
metal. We note that the thermopower of a cobaltate Na$_x$CoO$_2$
is remarkably enhanced due to the electronic spin
entropy\cite{NCO}. Thus we suggest that the anomalous thermopower
term might have a magnetic origin. Careful studies on the d.c.
magnetic susceptibility have found that the normal state magnetic
susceptibility shows indeed a dome-like doping dependence in
F-doped LaFeAsO$_{1-x}$F$_x$ system\cite{Chi}. This susceptibility
enhancement could be associated with spin fluctuations. Therefore
it was proposed that the spin fluctuations may play an important
role in the superconducting mechanism. However, the iron arsenide
system has very different nature in electronic state compared to
the sodium cobaltate system. In sodium cobaltate system, a strong
electron correlation picture is necessary to describe electronic
transport properties. The observation of suppression of
thermopower by magnetic field suggested a large spin entropy term
in thermopower. In contrast, the parent compounds LnFeAsO in the
iron arsenide system are not Mott insulators, and band
calculations\cite{Singh,wnl,Tesanovic,Xu} and transport property
measurements have suggested that the 3d electrons in this system
have itinerant nature. Therefore, the enhanced thermopower might
not from the spin entropy although we argue that it might have a
magnetic origin. How the spin fluctuations play an important role
in the electronic transport need further studies. If both the
enhanced thermopower and the enhanced susceptibility in the
superconducting window have indeed the common origin, the magnetic
fluctuations should also play an important role in the mechanism
of superconductivity.

\begin{figure}
\includegraphics[width=8cm]{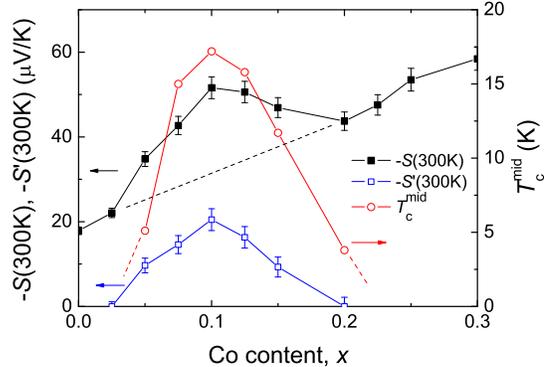}
\caption {(Color online) Doping dependence of room-temperature
thermopower, $S$(300), for SmFe$_{1-x}$Co$_{x}$AsO samples. The
superconducting transition temperature $T_{c}^{mid}$ is also shown
for comparison. The dashed line indicates the background term to
the thermopower. $S'$(300K) is the abnormally enhanced term, equal
to $S$(300K) subtracting the background normal term. See text for
detail.}
\end{figure}

\section{\label{sec:level1}Conclusion}

In conclusion, superconductivity and normal state transport
properties of Co doped SmFe$_{1-x}$Co$_{x}$AsO system exhibit
systematic variations with Co content. The SDW order is quickly
suppressed by Co doping, and superconductivity emerges as $x$
$\geq$ 0.05. Meanwhile there is a crossover from metal to
insulator in the normal state resistivity at low temperature. A
phase diagram is derived based on the transport measurements and a
dome-like $T_c$ versus $x$ curve is established. The maximum of
$T_c^{mid}$ is 17.2 K at the optimally doping level $x$ $\sim$
0.1. Furthermore, thermopower increases with Co doping, also shows
a maximum at $x \sim$ 0.10, and then decreases slightly with
decreasing $T_c$. After subtracting the background normal term,
the anomalous term of the room temperature thermopower,
$S'$(300K), also shows a dome-like doping dependence as
$T_{c}^{mid}$($x$) does. A close correlation between $T_c^{mid}$
and $S'$(300K) is proposed. This correlation may be associated
with the mechanism of superconductivity.

\begin{acknowledgments}
This work is supported by the National Science Foundation of
China, National Basic Research Program of China (No.2006CB601003
and 2007CB925001), and PCSIRT of the Ministry of Education of
China (No. IRT0754).
\end{acknowledgments}

Note added. - At completion of this work we became aware of one
paper by Y. Qi \emph{et al.} which reported Co-doping induced
superconductivity in SmFe$_{1-x}$Co$_x$AsO\cite{SmCo}. Only two Co
doping concentrations were investigated, and $T_c^{mid}$ of about
14.2 K was observed at $x$ = 0.10 in their report. Their result is
consistent with this paper.

\end{document}